\documentclass[showpacs, twocolumn]{revtex4-1}
\usepackage{graphicx}
\usepackage{amsmath}
\usepackage{appendix}
\begin{document}

\title{ Breathing modes of repulsive polarons in Bose-Bose mixtures }

\author{Abdel\^{a}ali Boudjem\^{a}a, Nadia Guebli, Mohammed Sekmane and Sofyan Khlifa-Karfa}
\affiliation{Department of Physics,  Faculty of Exact Sciences and Informatics, Hassiba Benbouali University of Chlef P.O. Box 78, 02000, Ouled Fares, Chlef, Algeria.}
\email {a.boudjemaa@univ-chlef.dz}

%\date{\today}

\begin{abstract}

We consider impurity atoms embedded in a two-component Bose-Einstein condensate in a quasi-one dimensional regime.
We study the effects of repulsive coupling between the impurities and Bose species on the equilibrium of the system for both miscible and immiscible mixtures
by numerically solving the underlying coupled Gross-Pitaevskii equations.  
Our results reveal  that the presence of impurities may lead to a miscible-immiscible  phase transition 
due to the interaction of the impurities and the two condensates. 
Within the realm of the Bogoliubov-de Gennes equations we calculate the quantum fluctuations due to the different types of interactions.
The breathing modes and the time evolution of harmonically trapped impurities in both homogeneous and inhomogeneous binary condensates
are deeply discussed in the miscible case using variational and numerical means. 
We show in particular that the self-trapping, the miscibility and the inhomogeneity of the trapped Bose mixture may strongly modify 
the low-lying excitations and the dynamical properties of impurities.  
The presence of phonons in the homogeneous Bose mixture gives rise to the damping of breathing oscillations of impurities width.

\end{abstract}

%\pacs{03.75.Hh, 67.60.Bc, 03.75.Mn, 67.85.Bc} 

\maketitle

\section{Introduction} \label{Intro}

The experimental realization of ultracold atomic gases provides a powerful platform for exploring many interesting
problems in many-body physics, in particular the properties of the Bose polaron problem which consists of an impurity interacts with  Bose-Einstein Condensate (BEC).
The polaronic effects are important to understand a wide range of phenomena such as transport properties of metals, semiconductors,  ionic crystals,
BEC \cite{Gersh} and DNA and proteins \cite{Gutz}. 
The dressing of a particle by a bosonic reservoir plays a crucial role in many other systems, such as ${}^3$He-${}^4$He mixtures \cite{Baym} 
and high temperature superconductors \cite{Dag}.
The observation of Bose polarons has been reported in many experiments \cite{Chik, Pal, Cat, Koh, Spe, Cat1,Scel, Fuk, Jor}.

From the theoretical side, the Bose polaron problem has aroused intense interest \cite{Dev, Grud}.
The Bogoliubov-Fr\"ohlich Hamiltonian \cite {Brud1, Temp, Cast, Cast1,  Shash, Grus, Vlt, Shcha, Kain1} 
has been succesfully used to describe the ground state properties of Bose polarons in the weak coupling regime.
Analytical and numerical studies of beyond-Fr\"ohlich polarons have been reported in Refs \cite {Tim1, THJ,Tim2, Blum, Brud, Jian, Ras, Jesp, Rath, Pen, Pars,  Grus1}
employing different techniques.
At finite temperatures, the properties of polarons have been addressed using the time-dependent-Hartree-Fock-Bogoliubov (TDHFB) theory \cite {Boudj,Boudj1, Boudj2, Boudj3}, 
diagrammatic approach \cite {Sun, Levin, NE}, and perturbative theory  \cite{Pastu}. 
Most recently, we have examined effects of quantum fluctuations on the dynamics of dipolar Bose polarons \cite{Guebli}. 

The above studies inspected only the case of impurities interact with a medium consisting of a single-component BEC. 
However, to the best of our knowledge, the physics of polarons in Bose-Bose mixtures i.e. impurity atoms immersed in a binary BEC, 
has not been addressed except the work of Ref.\cite{Compa}. 
Polarons in binary condensates allow us to understand, in useful manner, the intriguing coupling between the impurities and the two condensates
in the phenomenon of phase separation. 
Bose-Bose mixtures have attracted enormous interest in recent times due to their exceptional control of the inter- and intra-component interactions. 
These structures play a crucial role for observing new states of matter such as the droplet phase \cite{Petrov, Cab,Semg,Errico, Boudj12}.  
The phase separation and the miscible-immiscible phase transition are most important features of binary Bose mixtures.
The mean-field theory predicts that the mixture can be miscible (mixing between the two species) or immiscible (phase separation) 
depending on whether the miscibility parameter $ \Delta=g_1 g_2/g_{12}^2$ is larger or smaller than one, 
where ($g_1,g_2$) and $g_{12}$ are the bosonic intraspecies and interspecies interaction strengths, respectively. 
The phase separation can be affected by the strength of the interspecies interaction (see e.g. Ref.\cite{Wiem}) and thermal fluctuations \cite{Arko, Boudj00}.

The aim of this paper is to systematically investigate the static and dynamical properties of repulsive polarons in 
homogeneous and inhomogeneous Bose-Bose mixtures at zero temperature in a quasi-one-dimensional (1D) geometry. 
We do this in the spirit of Ref.\cite{THJ} where numerical and variational techniques were applied to model the dynamics of polarons in a single BEC.
We analyze the density profiles of the two condensates and the impurities by numerically solving 
the underlying coupled Gross-Pitaevskii (GP) equations for both miscible and immiscible mixtures.
We demonstrate that the mixture undergoes a transition from miscible to immiscible phase in the presence of the impurities due to the BEC-impurity interactions.
Such a transition is characterized by a change in the binding energy of the impurities with their host and a decrease in the condensate depletion \cite{Tom}.
This latter is calculated by solving the Bogoliubov-de Gennes (BdG) equations corresponding to the coupled GP equations.
Furthermore, we determine the effective potential and the breathing modes of the impurities for both homogeneous and trapped miscible mixtures. 
Our study is based on a variational scheme in the framework of the GP regime.
We show that the impurities become trapped in the localized spatial deformation of the condensates. 
%In harmonically trapped case we find that the inhomogeneity and the impurities bath may enhance the localization process.
It is found that the inhomogeneity,  Bose-impurity couplings and the miscibility of the mixture may strongly
affect the breathing modes and the motion of the impurities.
Similar to the single BEC, we find that the breathing oscillations of the impurities are damped due to the phonon-impurity interactions.
The analytical expressions obtained from the variational ansatz are checked by a direct numerical simulation of the coupled GP equations.

The remainder of this paper is structured as follows. 
In Sec.\ref{Model}, we present the basic ingredients of the formalism that describes the behavior of polarons in Bose-Bose mixtures. 
Section \ref{NR} presents the obtained numerical results of this model. 
The equilibrium density distributions of the two condensates and of the impurity component are analyzed for both miscible and immiscible mixtures.
In addition, we look at how the interspecies BEC-BEC and BEC-impurity interactions affect the binding energy of the system.
The total quantum depletion is computed by a direct numerical simulation of the BdG equations.
In Sec.\ref{BM} we report on an investigation of the breathing modes of harmonically trapped impurities inside homogeneous and inhomogeneous
Bose mixtures in the miscible case. To this end, we use  variational and numerical methods.
We quantitatively discuss the role of the inhomogeneity, Bose-impurity couplings and the miscibility parameter in the breathing oscillations  of the impurities. 
Section \ref{DIH} is devoted to the time evolution of the impurities width.
In Sec.\ref{concl} we conclude and discuss future work.

\section {Formalism} \label{Model}

\subsection{Coupled Gross-Pitaevskii equations}

We consider few impurity atoms of mass  $m_I$  embadded in a two-component BEC  with the atomic mass $m_j$ in a quasi-1D harmonic confinement 
for repulsive impurity-BEC couplings. 
In the quasi-1D geometry, the scattering lengths characterizing the intra-species interactions are obtained from the three-dimensional ones via
$g_j =2\hbar\omega_{j\perp} a_{j}$, where $j=1,2$ is the species label, $a_{j}$ is the boson-boson $s$-wave scattering length, 
and $\omega_{j\perp}$ is transverse trapping frequency which should be much larger than the longitudinal  trapping frequency 
$\omega_{jx}$ i.e. $\omega_{jx}/\omega_{j\perp}\ll1$.  The quasi-1D configuration requires also that 
the transverse frequency should be much larger of the chemical potential. Here the impurities are supposed to be pinned in the center of the trap.  
The regime of a weakly interacting mixture gas requires the correlation length of both condensates $\xi_{cj}= \hbar/ \sqrt{m_jn_jg_j}$ 
to be much larger than the mean interparticle separation $1/n_j$ \cite{Shly1} and the correlation length of the impurities $\xi_I= \hbar/ \sqrt{m_I n_I g_I}$ 
to be much larger than $1/n_I$. In such a case the fluctuations and depletion are small, and therefore, 
the dynamics of the system is governed by the coupled GP equations
\begin{align} 
i\hbar \dot{\Phi}_j & = \bigg( h_j^{sp}+g_j n_j + g_{12}  n_{3-j} + g_{Ij} n_I\bigg)\Phi_j ,  \label{T:DH1} \\  
i\hbar \dot{\Phi}_I & = \bigg( h_I^{sp}+\sum_{j=1}^2 g_{Ij} n_j\bigg)\Phi_I,  \label{T:DH2}  
\end{align}
where $n_j(x)=|\Phi_j(x)|^2$ is the density of each condensate and $n_I(x)=|\Phi_I(x)|^2$ is the density of the impurities,
$\Phi_j(x)= \hat\psi_j(x)- \hat{\bar \psi}_j(x)$ is the wavefunction of each condensate, 
$\hat\Psi_j $ and  $\hat{\bar \psi}_j$ are, respectively the total boson field operator, and the noncondensed part of the field operator,
$\Phi_I(x)= \hat\psi_I(x)- \hat{\bar \psi}_I(x)$ is the impurities wavefunction, 
$\hat\Psi_I $ and  $\hat{\bar \psi}_I$ are, respectively the total impurities field operator and the impurities fluctuations field operator. 
The single particle Hamiltonian for the condensates and the impurities are defined, respectively by $h_j^{sp}=-(\displaystyle\hbar^2/\displaystyle 2m_j) \nabla^2 + V_j$ and 
$h_I^{sp}=-(\displaystyle\hbar^2/\displaystyle 2m_I) \nabla^2 + V_I$, where $V_j(x)$ and $V_I(x)$ are respectively, the Bose mixture and the impurity trapping potentials.
%The coefficients $g_{Bj}=(4\pi \hbar^2/m_{Bj}) a_{Bj}$ and $g_{B12}=g_{B21}= 2\pi \hbar^2 (m_{B1}^{-1}+m_{B2}^{-1}) a_{B12}$ with 
%$a_{Bj}$ and $a_{B12}$ being the bosonic intraspecies and interspecies coupling constants, respectively.
The coefficient $g_{Ij}$ stands for the impurity-boson interaction which can be determined numerically \cite{Cat1, Compa, Ling}.
Setting  $g_{Ij}=0$, Eqs.(\ref{T:DH1}) reduce to the coupled GP equations describing  
the dynamics of Bose mixtures at zero temperature, while Eq.(\ref{T:DH2}) reduces to the usual Schr\"odinger equation.
For $g_{12} =0$, one reproduces the standard GP equations employed for single Bose polarons. 
The nonlinear term proportional to $g_I$ (the impurity interaction strength) is neglected since the number of impurity atoms 
is assumed to be extremely small \cite {Tim1, THJ, Boudj}. 

\begin{figure*}
\begin{center}
\includegraphics[scale=0.5]{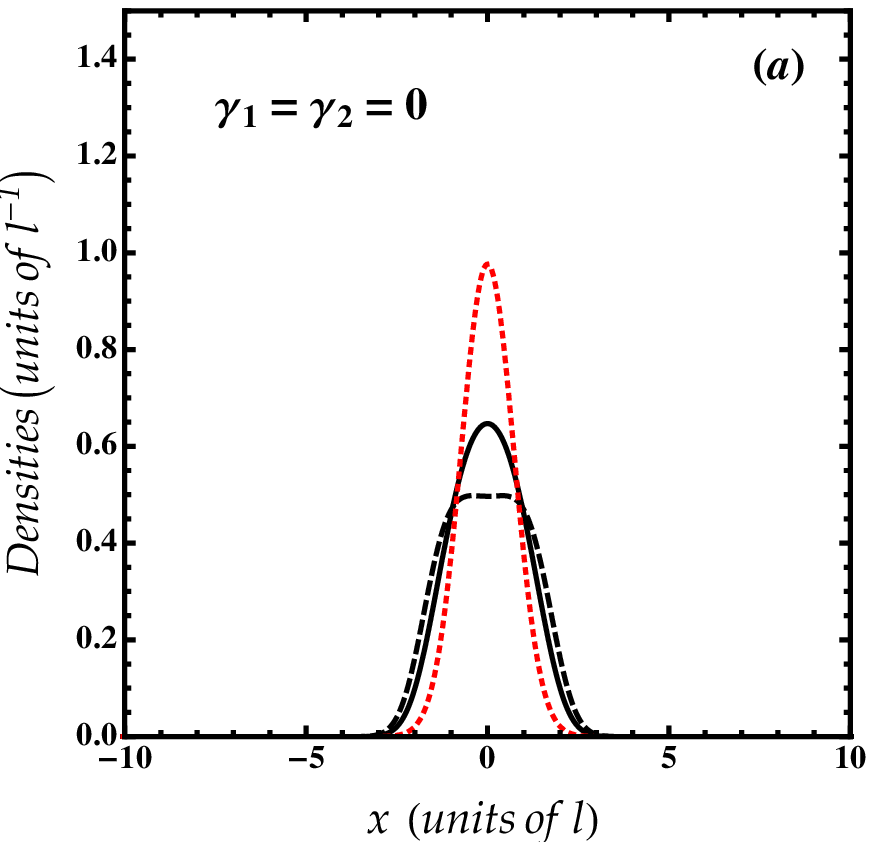} 
\includegraphics[scale=0.5]{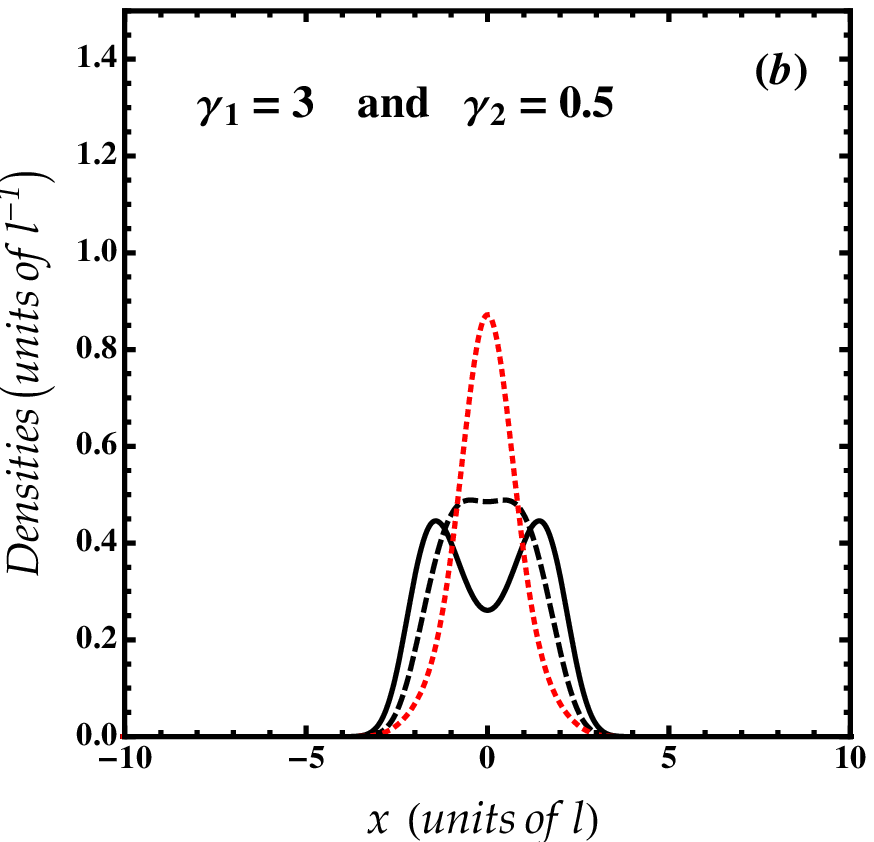} 
\includegraphics[scale=0.5]{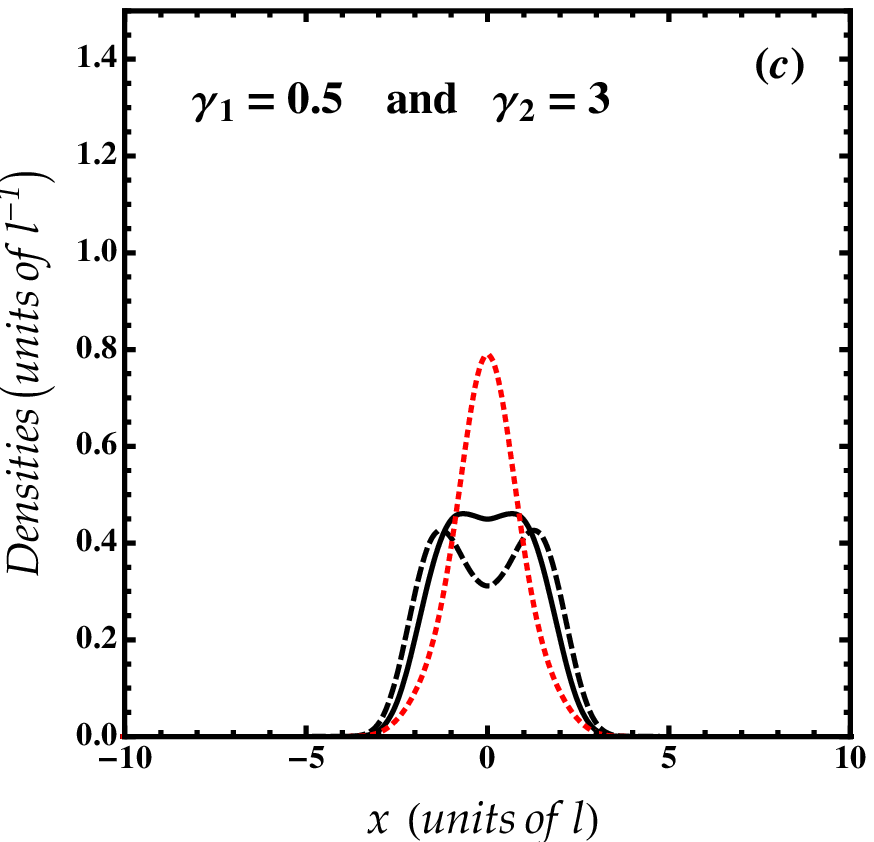} 
\includegraphics[scale=0.5]{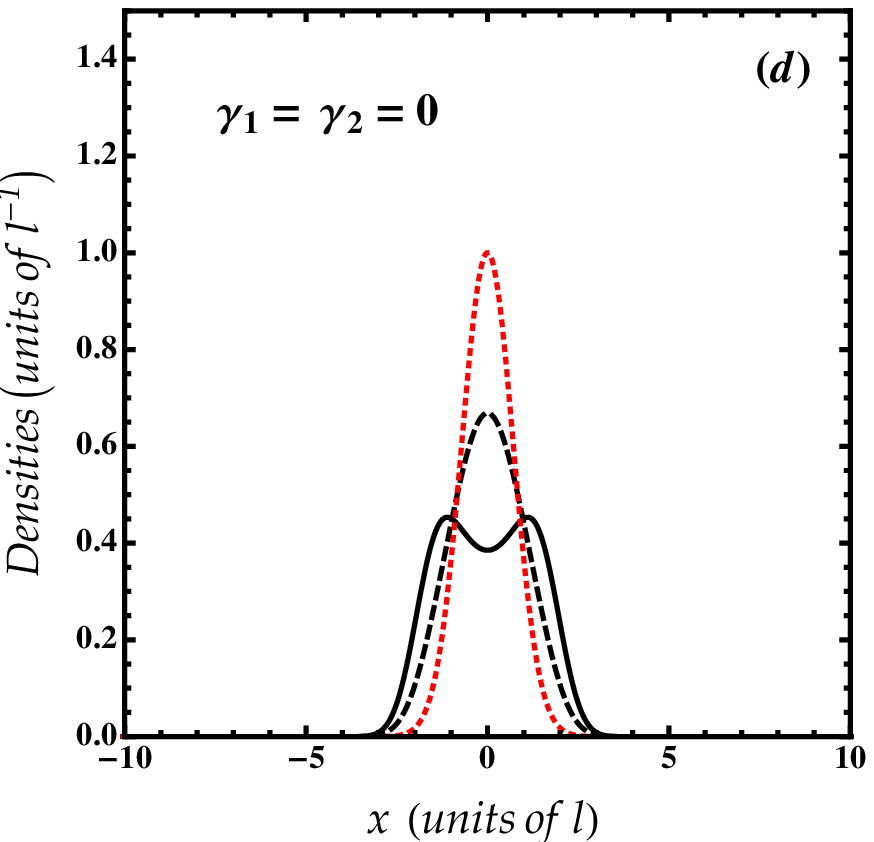} 
\includegraphics[scale=0.5]{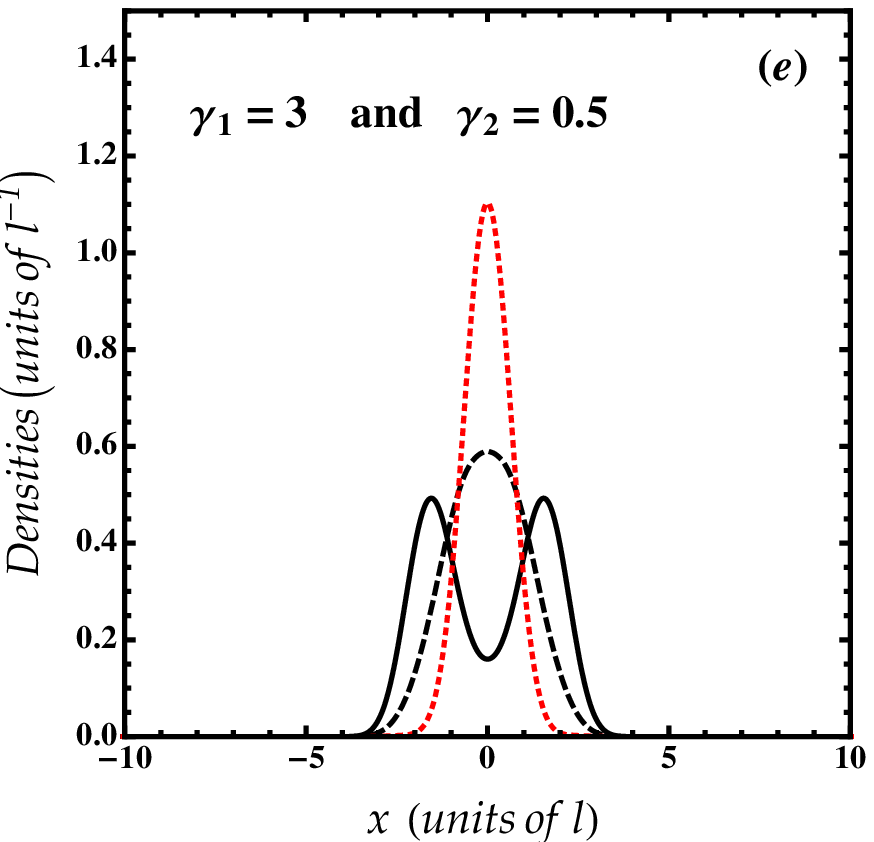} 
\includegraphics[scale=0.5]{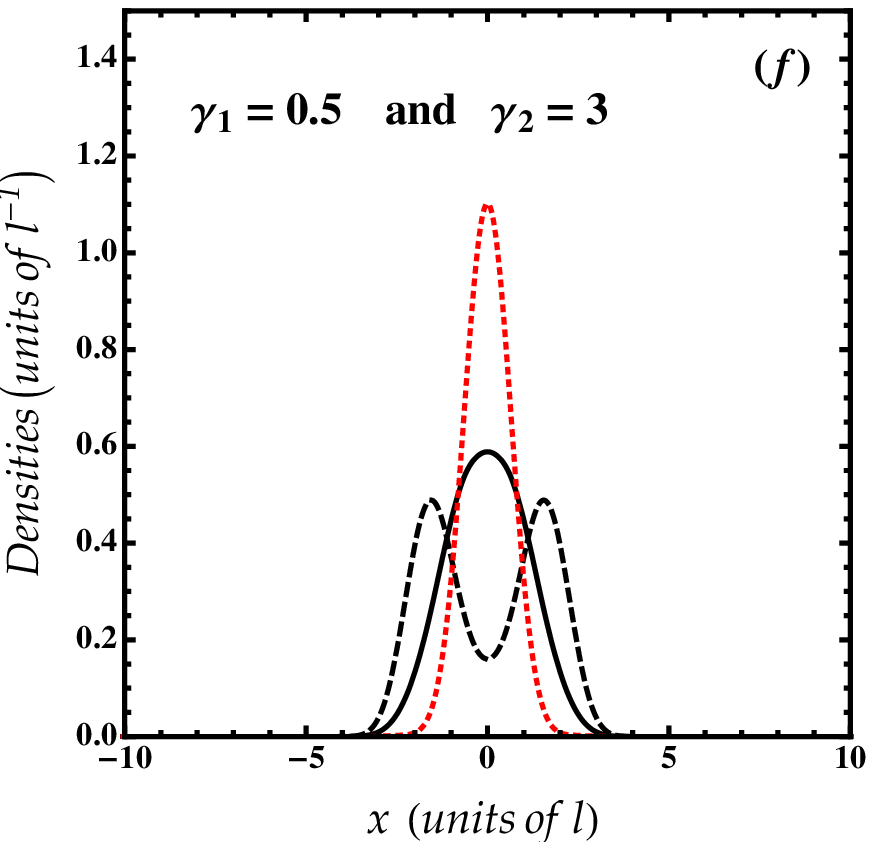} 
  \caption{ (a)-(c) Density profiles of impurities (red dotted lines),  BEC 1 (black solid lines), and BEC 2 (black dashed lines) for different values
 of $ \gamma_1 $ and $ \gamma_2 $, and $\Delta=1.9$ (miscible Bose-Bose mixture). (d)-(f) The same but for  $\Delta=0.9$ (immiscible mixture).
The density of the impurities has been amplified by 10 times for clarity.}
    \label{DF}
 \end{center}  
\end{figure*}

The static equilibrium equations can be found using 
$\Phi_j (x,t)= \Phi_j (x) \exp(-i \mu_j t/\hbar)$, and $\Phi_I (x,t)= \Phi_I (x) \exp(-i \mu_I t/\hbar)$,
where $\mu_j$ are chemical potentials related with bosonic components and $\mu_I$ stands for the chemical potential of impurity atoms. 
This gives
\begin{align} 
\mu_j \Phi_j &= \bigg( h_j^{sp}+g_j n_j + g_{12}  n_{3-j} + g_{Ij} n_I\bigg)\Phi_j,  \label{sgpe1}  \\
\mu_I \Phi_I & = \bigg( h_I^{sp}+\sum_{j=1}^2 g_{Ij} n_j\bigg)\Phi_I.    \label{sgpe2}
\end{align} 
Note that the chemical potential of each species can be determined from the normalization conditions:
$N_j=\int n_j d x $ is the single condensate total number of particles, and $N_I=\int n_I d x$, 
is the number of impurities.

The energy functional corresponding to (\ref{T:DH1}) and  (\ref{T:DH2}) reads
\begin{align}  \label{egy}
E&= \sum_{j=1}^2 \int d x \bigg[ \frac{\hbar^2}{2m_j} |\nabla \Phi_j |^2+ V_j n_j+\frac{1}{2} g_j n_j^2 \\
&+\frac{1} {2} g_{3-j} n_{3-j}^2 + g_{12} n_j n_{3-j}+ g_{Ij} n_j n_I \bigg]\nonumber \\
&+ \int dx \bigg( \frac{\hbar^2}{2m_I} |\nabla \Phi_I |^2+ V_I n_I \bigg)\nonumber.
\end{align}
The minimum of this energy in terms of the mean distance between the condensates and impurities $d_j = \int d x x  \left [n_j (x, t) -n_I (x,t) \right]$
will give insights into the binding mechanism of the impurities state. For simplicity,  we assume that the impurities are 
placed equidistant from the two condensates.
One can expect that the inter-BEC ($g_{12}$) and BEC-impurity ($g_{Ij}$) interactions may alter the binding energy of the impurities with their host.

%The time dependence of the mean separation between the two condensates is given  by
%\begin{equation}  \label{MdS}
%d_j = \int d x x  \left [n_j (x, t) -n_I (x,t) \right],
%\end{equation}

\subsection{Bogoliubov-de-Gennes equations}

The study of Bogoliubov excitations and quantum fluctuations amounts to solving the so-called BdG equations.
To this end, we linearize Eqs.(\ref{T:DH1}) and (\ref{T:DH2}) around static solutions as
$\Phi_j=\Phi_{0j}+\delta \Phi_j$ and $\Phi_I=\Phi_{0I}+\delta \Phi_I$,
where $\delta \Phi_j / \Phi_{0j}\ll 1$ and $\delta \Phi_I/ \Phi_{0I}\ll 1$  \cite{Boudj1,Tim1,Boudj12}. 
Here we assume that the presence of the impurity atoms does not perturb the condensate significantly \cite{Tim1}.
This yields
\begin{align} 
i\hbar \delta  \dot \Phi_j & =\bigg( h_j^{sp}+2g_j |\Phi_{0j}|^2 + g_{12}  |\Phi_{03-j}|^2+ g_{Ij} |\Phi_{0I}|^2\bigg)\delta \Phi_j \nonumber   \\ 
&+ g_j \Phi_{0j}^2\delta \Phi_j^* +g_{12} \Phi_{0j} \Phi_{03-j} \delta \Phi_{3-j}^* +g_{12} \Phi_{0j} \Phi_{03-j}^* \delta \Phi_{3-j} \nonumber   \\ 
&+ \Phi_{0j} \Phi_{0I} \delta \Phi_I^*+g_{Ij} \Phi_{0j} \Phi_{0I}^* \delta \Phi_I, \label {RPA1}  
\end{align}
and 
\begin{align} 
i\hbar \delta  \dot \Phi_I & =\bigg( h_I^{sp}+\sum_{j=1}^2 g_{Ij}|\Phi_{0j}|^2\bigg)\delta \Phi_I +\sum_{j=1}^2 g_{Ij} \Phi_{0I}  \Phi_{0j} \delta \Phi_j^*  \nonumber   \\ 
&+ \sum_{j=1}^2 g_{Ij} \Phi_{0I}  \Phi_{0j}^*\delta \Phi_j. \label {RPA2}  
\end{align}
The coupled BdG equations consist of writing the field fluctuations associated with the two condensates and the impurities in the form :
$\delta \Phi_j ({\bf r},t)= u_{jk}  e^{i {\bf k \cdot r}-i\varepsilon_k t/\hbar}+v_{jk} e^{i {\bf k \cdot r}+i\varepsilon_k t/\hbar}$, 
and $\delta \Phi_I ({\bf r},t)= u_{Ik}  e^{i {\bf k \cdot r}-i\varepsilon_k t/\hbar}+v_{Ik} e^{i {\bf k \cdot r}+i\varepsilon_k t/\hbar}$, 
where $u_{kj}$ and $v_{kj}$ are the Bogoliubov quasiparticle amplitudes and $\varepsilon_k$ is the Bogoliubov excitation energy. 
The resulting BdG equations yield a generalized excitations spectrum due to the interspecies interactions and the impurities corrections.
The obtained spectrum exhibits numerous quasiparticle properties of the attractive, repulsive and molecular branches.
In the absence of impurities, the excitations spectrum reduces to that obtained using for example the TDHFB-RPA theory \cite{Boudj00, Boudj12} and found to be composed of two branches: 
upper branch $\omega_+$ and  lower branch $\omega_-$. This latter becomes complex for  $\Delta <1$.

The depletion of each component induced by the intra- and interspecies interactions between condensed bosons is defined as:
$\tilde n_j (x) =\langle \hat{\bar \psi}_j^\dagger (x)  \hat{\bar \psi}_j (x)  \rangle$. 
At zero temperature, it can be written in terms of the Bogoliubov quasiparticle amplitudes as:
 \begin{equation}\label {Cdep}
\tilde n_j (x) = \sum_k v_{jk}^2(x).
\end{equation}
However, the depletion caused by the BEC-impurity interactions is given by $\tilde n_I (x) =\langle \hat{\bar \psi}_I^\dagger (x) \hat{\bar \psi}_I(x) \rangle= \sum_k v_{Ik}^2(x)$.
The total depletion is defined as $\tilde n=\sum_{j=1}^2\tilde n_j+\tilde n_I$.\\
In this paper we inspect in particular the impacts of impurities on the condensate depletion and on the phase separation of binary BEC.
In the miscible phase, the condensate depletion increases with interactions while in the immisicble regime it decreases with interactions \cite{Tom}.
Therefore, the transition from a miscible to an immiscible phase is marked by a decay depletion.

%It is well known that in the absence of impurities, the mixture is fully coherent i.e. $g_j^{(1)}$ tends to $n_j$ at large distances in both miscible and immiscible cases. 
%However, the presence of impurities in the system may lead to modify the form of the first-order correlation function
%which can be considered as a clear signature of a miscible-immiscible phase transition.

\section{Numerical results} \label{NR}

%\subsection{Density profiles}

To be quantitative,  we consider few impurities of ${}^{41}$K atoms in a bath of a mixture of two hyperfine states of ${}^{87}$Rb atoms. 
Our simulations are performed for a Bose mixture of equal  masses, harmonic frequencies and numbers of atoms.
The parameters are set to: $N_1=N_2=300$ atoms, $N_I=5$ atoms for impurity, 
$g_1 \simeq2.08 \times 10^{-37}$ J.m,  $g_2 \simeq 1.99 \times 10^{-37}$ J.m  \cite{Cat1,Compa}.
The interspecies scattering length $g_={12}$ can be adjusted by means of a Feshbach resonance to reach miscible/immiscible mixtures.
The transverse and the longitudinal trapping frequencies of the two condensates and the impurities are given, respectively as: $\omega_{1 \perp}=\omega_{2\perp}=2\pi \times 34$ kHz,  
$\omega_{1x}=\omega_{2x}=2\pi \times 62$ Hz \cite{Cat1}, $\omega_{I \perp}=2\pi \times 50$ kHz,  and $\omega_{I x}=2\pi \times 90$ Hz. 
Lengths and energies are expressed in terms of  $l=\sqrt{\hbar/ m\omega_{x}}$ and $\hbar \omega_x$, respectively.
We introduce the dimensionless parameters $\gamma_1=g_{I1}/g_1$ and $\gamma_2=g_{I2}/g_2$
describe the relative strengths of interactions.

We then solve Eqs.(\ref{sgpe1}) and (\ref{sgpe2}) numerically for both miscible and immiscible mixtures.
The results  are shown in Fig.\ref{DF}. 
We see that when the two condensates and the impurities are decoupled i.e. $\gamma_1 = \gamma_2 =0$, the mixture remains miscible (see Fig.\ref{DF}.a). 
Moreover, for $\gamma_1 > \gamma_2$, the impurities which develop a higher peak density and a narrower width, 
create a localized spatial deformation in BEC1 and get confined inside to it, while BEC2 remains intact (see Fig.\ref{DF}.b).
When $\gamma_2 > \gamma_1$, the situation is quite the opposite notably BEC2 becomes deformed while BEC1 is moderately distorted by impurities (see Fig.\ref{DF}.c). 
This feature has never been observed yet in the literature, and can serve as a signature of the miscible-immiscible phase transition.  
From Fig.\ref{DF}.a-c, we see also that the impurities density distribution is slightly reducing with the BEC-impurity interactions.

Figure.\ref{DF}.d shows that in the case of an immiscible mixture, for $\gamma_1 = \gamma_2 =0 $, the system preserves its form i.e. 
BEC1 is pushed towards the outer part forming a shell structure around BEC2 \cite{Boudj00, Prouk}. 
For $\gamma_1 > \gamma_2 $, BEC1 is pushed out of the center, but the impurities come near the center by being pushed by the surrounding bosons, 
while BEC2 keeps its form (see Fig.\ref{DF}.e).
For $\gamma_2 > \gamma_1$, the situation is quite different where BEC2 is profoundly distorted leading to a pronounced spatial phase separation.
Whereas, BEC1 keeps its robustness (see Fig.\ref{DF}.f). This is most probably due to the interplay of the BEC-BEC and BEC-impurity couplings. 
In contrast to the miscible case, the density of impurities slightly increases with the BEC-impurity coupling as is shown in Fig.\ref{DF}.d-f.

\begin{figure}
\includegraphics[scale=0.46]{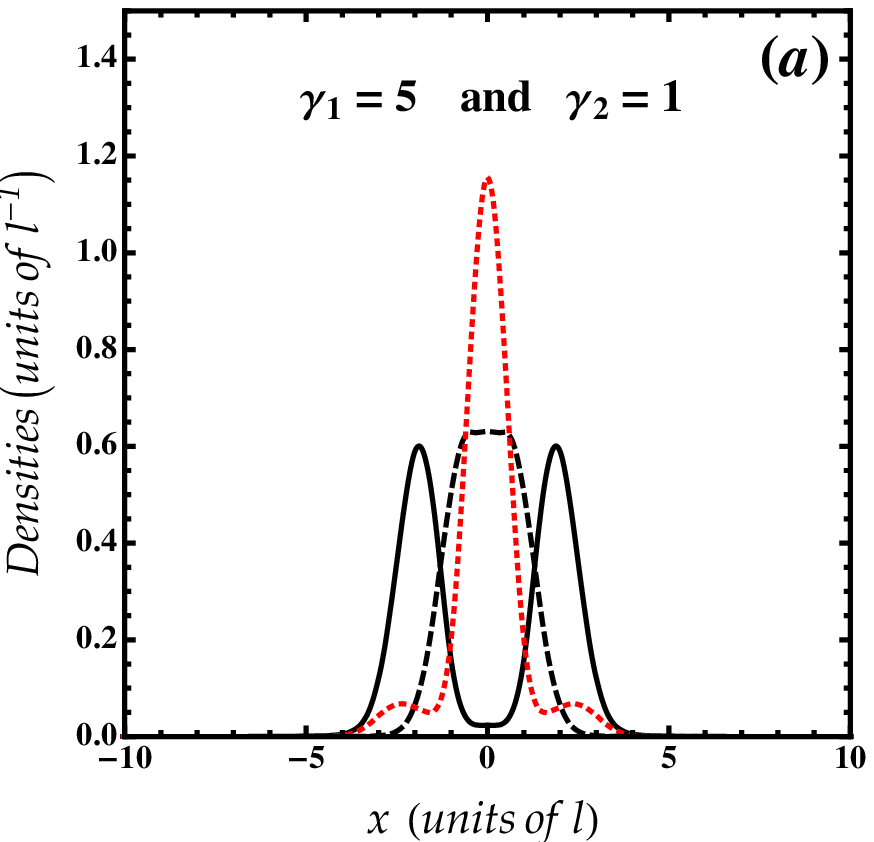} 
\includegraphics[scale=0.46]{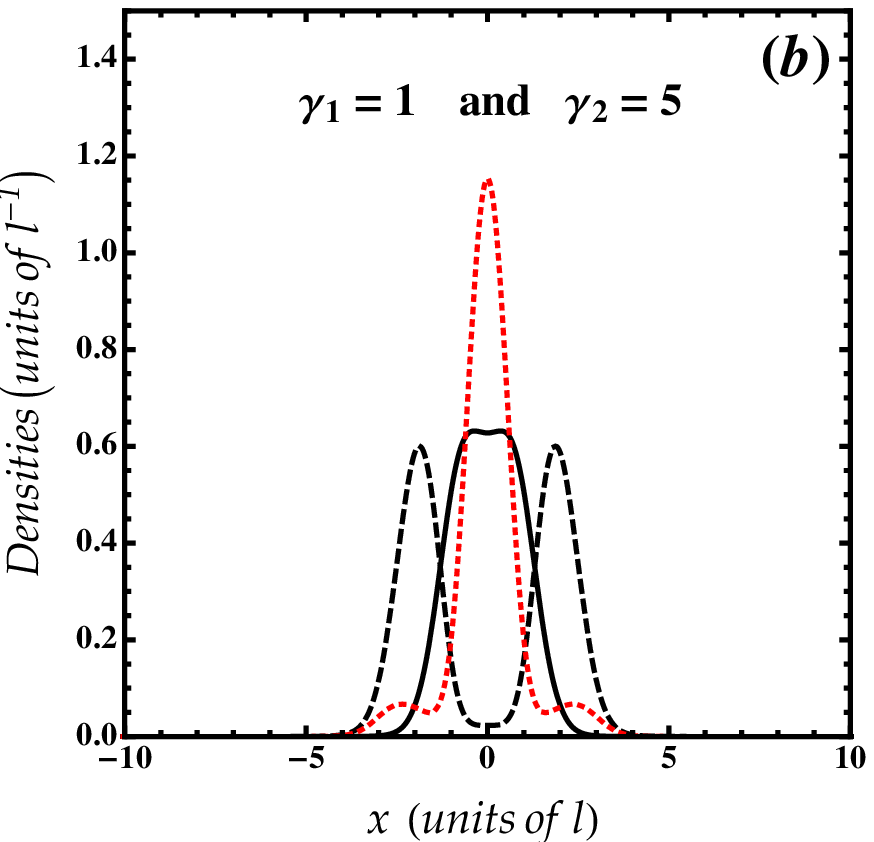} 
  \caption{ Density profiles of impurities strong BEC-impurity coupling for $\Delta=1.9$ (a) and $\Delta=0.9$ (b).
Parameters are the same as in Fig.\ref{DF}.}
    \label{bind}
 \end{figure}

For strong repulsive BEC-impurity interactions,  the impurity-induced dip, becomes deeper and deeper until 
one of the two condensates breaks into two fragments, while the second component remains robust and get trapped entirely inside the first BEC (see Fig.\ref{bind}).
This behavior holds also for polarons in a single BEC \cite{Akr}. 
We observe also that the impurities develop a small structure at the borders.

\begin{figure}
\includegraphics[scale=0.46]{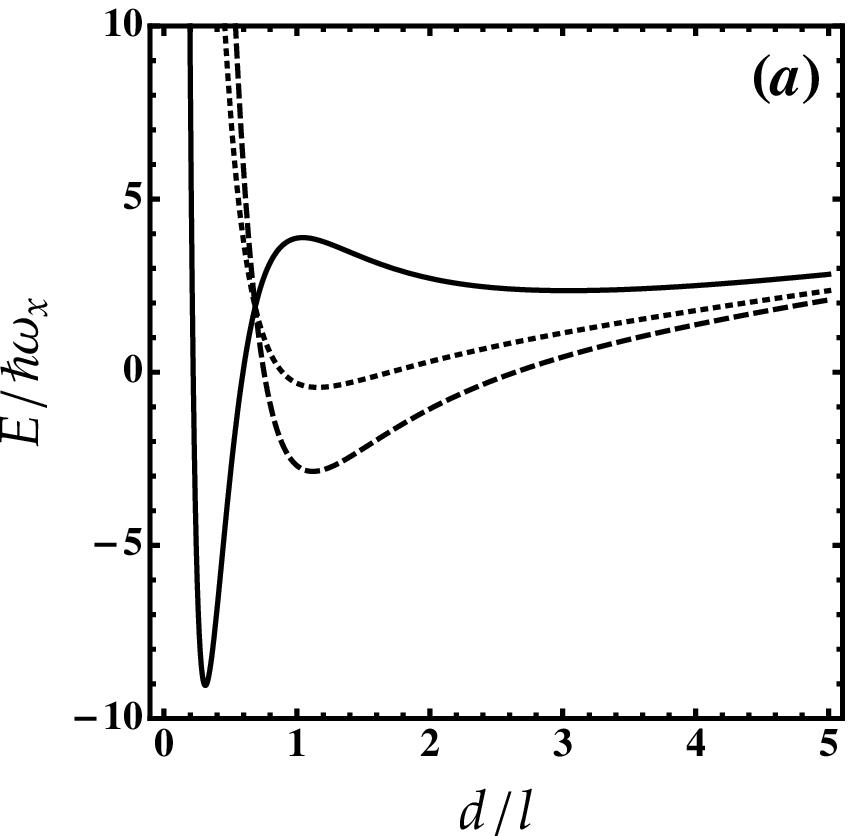} 
\includegraphics[scale=0.46]{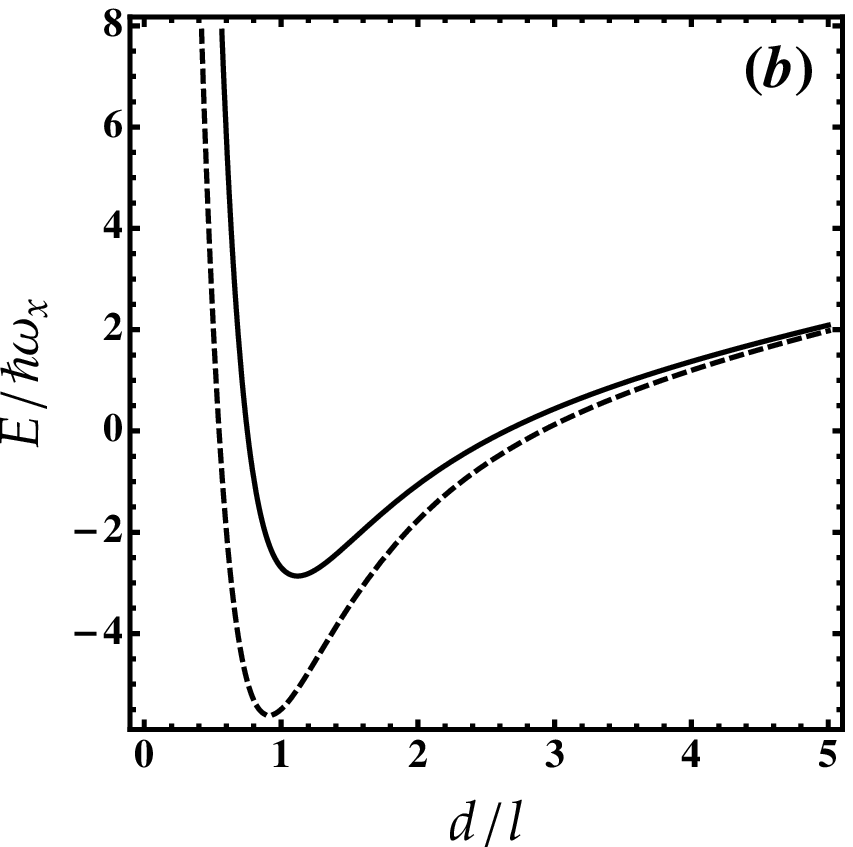} 
  \caption{ (a) Binding energy as a function of the mean distance $d$ for different values of interspecies interactions strength $g_{12}$. 
  Solid line: $g_{12} = 2.14 \times 10^{-37}$ J.m.  Dashed line:  $g_{12} = 1.67 \times 10^{-37}$ J.m. Dotted line:  $g_{12} = 0.64 \times 10^{-37}$ J.m.
Here the coefficients $g_1$ and $g_2$ are fixed.
 (b) Binding energy as a function of the mean distance $d$ for different values of $\gamma$ for $\Delta=1.5$. 
Solid line: $\gamma_1=0.5$ and  $\gamma_2=3$. Dashed line: $\gamma_1=1$ and  $\gamma_2=5$.
Parameters are the same as in Fig.\ref{DF}.}
    \label{SCQF}
 \end{figure}

We show now that this variation of the density profiles is correlated with the change in the binding energy of the system 
and hence, to the miscible-immiscible phase transition.
As illustrated in Fig.\ref{SCQF}.a, the binding energy is decreasing with the inter-BEC interaction strength $g_{12}$ indicating that the 
impurities start to delocalize. If $g_{12}$ approaches zero, they could be completely delocalized. 
In the immiscible case ($g_{12} = 2.14 \times 10^{-37}$ J.m or equivalently $\Delta=0.9$), the impurities are strongly bound in their bosonic mixture host
in particular at small distances.
Figure \ref{SCQF}.b depicts that for fixed $g_{12}$, the polarons has indeed nonzero binding energy in terms of the BEC-impuirty interactions $\gamma_j$. 
Quite striking is the depth of the minimum increases with $\gamma_j$ even in the miscible case a fact that strengthens the localization of the impurities in their bath.
This dramatic modification of $E$ tells us that a transition from the miscible to immiscible phase is occured.

%\subsection{Quantum fluctuations}

\begin{figure}
\includegraphics[scale=0.8]{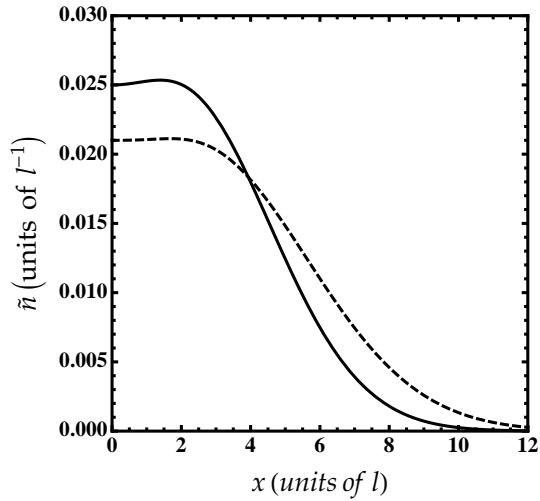} 
  \caption{ Quantum depletion due to the boson-boson  and boson-impurity interactions for $\Delta=1.9$.
Solid lines correspond to $\gamma_1=\gamma_2=0$.
Parameters are the same as in Fig.\ref{DF}.}
    \label{QF}
 \end{figure}

To gain additional insight into the miscible-immiscible phase transition, we analyze
the quantum depletion caused by both boson-boson and boson-impurity interactions.
To this end, we solve iteratively our BdG equations together with Eqs.(\ref{sgpe1}) and (\ref{sgpe2}).
Figure \ref{QF}.a shows that the presence of impurities in the two condensates leads to decrease the total depletion even for $\Delta >1$ 
results in a transition to immiscible phase. 
In such a case the BEC-BEC interaction becomes very weak giving rise to a small quantum depletion since the condensates separate from each othe.
Remarkably, the quantum depletion is around $2.5\%$ of the total gas density which highly justifies the use the coupled GP equations as is anticipated above.
For typical weakly interacting ultracold Bose gas experiments the depletion is $\sim 1\%$ \cite{Lop}.

%As illustrated in Fig.\ref{QF}.b, for $\gamma_1=\gamma_2=0$, $g_j^{(1)}(x)$ keeps its usual form and decays towards $n_j$ at large distances
%indicating the existence of the long-range order in both components. 
%We see also that the dilute component (BEC2) has a broaden maximum near at the center.
%Note that the coherence remains also perfect in the immiscible mixture ($\Delta=0.9$) as we have stated above.
%Quite striking is the narrow peak at the center of the first-order correlation function and its oscillating behavior at large distances which originate from the presence of the impurities.
%This dramatic modification of $g_j^{(1)}(x)$ tells us that a transition from the miscible to immiscible phase is occured.

\section{Breathing modes} \label{BM}

The focus in this section is upon the breathing modes of Bose mixture polarons in both homogeneous and inhomogeneous cases.
Our analysis is based on a variational method and numercial simulation of the BdG equations. 

Let us assume that our system to be in the Thomas-Fermi (TF) regime where the kinetic term associated with each condensate is negligible. Then Eqs.(\ref{sgpe1}) reduce to
\begin{align} \label{TF} 
n_j= &\frac{\Delta}{ (\Delta -1)} \bigg [ n_{0j}- \frac{g_{12}} {g_j} n_{0,3-j} \\
&- \left(\gamma_j-\frac{g_{12}} {g_j} \gamma_{3-j} \right) n_I  \bigg], \;\;\;\;\;\;   j=1,2 \nonumber
\end{align}
where $n_{0j}=(\mu_j-V_j)/g_j$ is the decoupled TF condensates density.
For $\gamma_j=0$,  the two condensates reduce to their "decoupled values", $[\Delta/ (\Delta -1)] ( n_{0j}- n_{0,3-j} g_{12}/g_j )$. 
For $\gamma_j \neq 0$, Eqs.(\ref {TF}) indicate that the BEC-impuirty interactions may lead to shift the two BEC densities  from their decoupled values.\\
The weak impurity-Bose mixture coupling requires the inequality: 
\begin{equation} \label{VC}
n_{0j}-  \frac {g_{12}} {g_j} n_{0,3-j}  \gg \left(\gamma_j-  \frac {g_{12}} {g_j} \gamma_{3-j} \right) n_I,
\end{equation}  
which means that the two BEC densities must be larger than the impurities density.
For mixtures having equal densities $n_{0j}= n_{0,3-j}=n_0$,  the condition (\ref{VC}) simplifies to
$(1- g_{12}/g_j) n_0 \gg \left(\gamma_j- \gamma_{3-j} g_{12}/g_j  \right) n_I$.

Introducing Eqs.(\ref{TF}) into Eq.(\ref{T:DH2}), one obtains the extended self-focussing nonlinear Schr\"odinger equation (NLSE)
\begin{align}\label{GPE}
i\hbar\dfrac{\partial\Phi_{I}}{\partial t}&=\Bigg \{-\dfrac{\hbar^{2}}{2m_I}\nabla^2+V_I+\Big(\dfrac{\Delta}{\Delta-1}\Big) \sum_{j=1}^2 g_{Ij}\\
&\times\bigg [ n_{0j} - \frac{g_{12}} {g_j} n_{0,3-j}- \left(\gamma_j-\frac{g_{12}} {g_j} \gamma_{3-j} \right) n_I  \bigg]\Bigg\}\Phi_I. \nonumber
\end{align}
This equation is appealing since it describes the dynamics of harmonically trapped impurity-Bose-Bose mixtures in terms of the miscibility parameter, $\Delta$. 
For $g_{12}=0$, it reduces to the habitual NLSE for a single BEC-impurity mixture.
In the absence of the impurities trapping potential ($V_I=0$), Eq.(\ref{GPE}) admits an almost exact solution for weakly localized impurities \cite{Brud,Boudj2,DAnd}. 
It is valid only for the miscible regime since the TF approximation is less satisfactory for phase separation.
From now on we consider polarons in a miscible mixture.

Equation (\ref{GPE}) can be solved using the following variational ansatz \cite{THJ,Guebli}
\begin{equation} \label{Gaus}
\Phi_{I}(x,t)=A(t)\exp\bigg[-\dfrac{x^2}{2\sigma^2(t)}-i\beta(t) x^2\bigg],
\end{equation}
where $ A(t) $ is the normalization constant, $ \beta(t) $ is the phase and $\sigma(t)$ is the width of the impurity.
The normalization condition $ \int_{-\infty}^{+\infty}\Phi_{I}^{*}\Phi_{I}d x=N_I $, yields $ A(t)=[N_I^2/\pi \sigma(t)^{2}]^{1/4} $.\\

The Lagrangian density corresponding to Eq.(\ref{GPE}) is given as
\begin{align}\label{LL}
{\cal L} &= \frac{i \hbar}{2} \left(\Phi_I  \frac{\partial \Phi^*_I}{\partial t}-\Phi_I^*  \frac{\partial \Phi_I}{\partial t}\right) +\frac{\hbar^{2}}{2m_{I}} |\nabla\Phi_I|^2\\
&-V_I |\Phi_I|^2-\Big(\dfrac{\Delta}{\Delta-1}\Big) \sum_{j=1}^2 g_{Ij} \bigg [ n_{0j} |\Phi_I|^2 \nonumber\\
&- \frac{g_{12}} {g_j} n_{0,3-j} |\Phi_I|^2+ \left(\gamma_j-\frac{g_{12}} {g_j} \gamma_{3-j} \right) |\Phi_I|^4 \bigg]. \nonumber
\end{align}
%We suppose that the  Bose mixture and the impurity are confined in a quasi-1D harmonic potential $V_{Bj}=m_j \omega_j^2 x^2/2$ and $V_I=m_I \omega_I^2 x^2/2$.
Then inserting the ansatz (\ref{Gaus}) into Eq.(\ref{LL}), obtain the Lagrangian $L=\int_{-\infty} ^{\infty} {\cal L} dx$, 
and derive the following Euler-Lagrange equations for the phase and the width of the impurity atoms
 \begin{equation}\label{Veq1}
\beta(t)=-\frac{m_{I}\dot{\sigma}(t)} {2\hbar\sigma(t)},
\end {equation}
and
\begin{equation}\label{Veq2}
m_{I}\ddot{\sigma}=-\dfrac{\partial \cal V (\sigma)} {\partial \sigma},
\end {equation}
where the effective potential $\cal V (\sigma)$ reads
\begin{align}\label{pot1}
 \cal V(\sigma)&=\dfrac{\hbar^{2}}{2m_I\sigma^{2}}+\frac{m_I}{2}\omega^{2}_{I}\sigma^{2} \\ 
&-\Big(\dfrac{\Delta}{\Delta-1}\Big) \sum_{j=1}^2  \bigg\{\dfrac{4 g_{Ij} N_I^{1/2}(\gamma_j- \gamma_{3-j} g_{12}/g_j )}{(2\pi)^{1/2}\sigma} \nonumber\\
&+2  \left(1+\frac{g_{12}}{g_j}\right) \mu_j
\Big[2\tilde{\Gamma}\Big(\dfrac{1}{2},\dfrac{R_j^2}{\sigma^2}\Big)-\dfrac{\sigma^2}{R_j^2}\tilde{\Gamma}\Big(\dfrac{3}{2},\dfrac{R_j^2}{\sigma^2}\Big)\Big]\bigg\}. \nonumber
\end{align}
where $ \Gamma(p)=\int_{0}^{+\infty} dx\exp^{-x}x^{p-1} $ is the Gamma function, and $R_j=\sqrt{2\mu_j/m_j\omega_j^2}$ is the TF radius of each BEC.
In the quasi-1D regime the inequality $ R_j \ll  l_j$ should be satisfied \cite{Peth, Boudj9}.
The normalized lower incomplete Gamma function is defined as $\tilde{\Gamma}(p,z)=[\Gamma(p)]^{-1}\int_{0}^{z}dx\exp^{-x}x^{p-1} $.
The first term on the right-hand-side in Eq.(\ref{pot1}) is the kinetic energy, and the following term stands for harmonic trapping of the impurity. 
The third term represents self-trapping effects and comes from the deformation of the mixture due to the host-impurity and host-host interactions. 
The last contribution describes the inhomogeneity in the mixture owing to its trapping potential. 
Equations (\ref{Veq1})-(\ref{pot1}) constitute a natural extention of those obtained in Ref.\cite{THJ} for a single BEC with impurity.

The low-lying excitations of the impurity are computed by expanding the width (\ref{Veq2}) around the equilibrium :
$\sigma =\sigma_0+\delta \sigma$, where $\sigma_0$ is the equilibrium width and $\delta \sigma/\sigma_0 \ll 1 $ 
which results in $\omega= \sqrt{\partial^2  {\cal V}/ \partial \sigma^2}|_{\sigma=\sigma_0}$.
In the phase separation regime when $\Delta \rightarrow 1$, the frequency $\omega \rightarrow \infty$ and hence, the system becomes unstable.

%\begin{widetext}
%\begin{align}\label{LLL1}
%L&=\dfrac{3N_{I}^{3/2}}{2}\hbar\dot{\beta}\sigma^{2}-\dfrac{3\hbar^{2}N_{I}^{3/2}}{2m_{I}}\Big(\dfrac{1}{2\sigma^{2}}+2\beta^{2}\sigma^{2}\Big)-\dfrac{3N_{I}^{3/2}}{4}m_{I}\omega^{2}_{I}\sigma^{2} 
%+N_{I}^{3}\Big(\dfrac{\Delta}{\Delta-1}\Big)\dfrac{(g_{1}\gamma_{I1}^{2}+g_{2}N_{I}^{3/2}\gamma_{I2}^{2}-2g_{12}\gamma_{I1}\gamma_{I2})}{(2\pi)^{3/2}\sigma^{3}} \nonumber\\
%&-N_{I}^{3/2}\Big(\dfrac{\Delta}{\Delta-1}\Big)(\gamma_{I1}\tilde{\mu_{1}}+\gamma_{I2}\tilde{\mu_{2}}) 
%\dfrac{3}{2}\Big[\dfrac{2}{3}\tilde{\Gamma}\Big(\dfrac{3}{2},\dfrac{\tilde{R}^{2}}{\tilde{\sigma}^{2}}\Big)
%-\dfrac{\tilde{\sigma}^{2}}{\tilde{R}^{2}}\tilde{\Gamma}\Big(\dfrac{5}{2},\dfrac{\tilde{R}^{2}}{\tilde{\sigma}^{2}}\Big)\Big].\nonumber
%\end{align}
%\end{widetext}
%Variational equations of motion can be derived using the standard Euler-Lagrange equations $\frac{\partial}{\partial t}\left(\frac{\partial L}{\partial\dot{q}}\right)-\frac{\partial L}{\partial q} $.

\subsection {Homogeneous mixture}

Let us start by considering impurity atoms inside a homogeneous mixture.
In such a situation, the decoupled condensates density becomes uniform and 
the confining potential (last term in Eq.(\ref{pot1})) can be ignored.

\begin{figure}
\begin{center}
\includegraphics[scale=0.45]{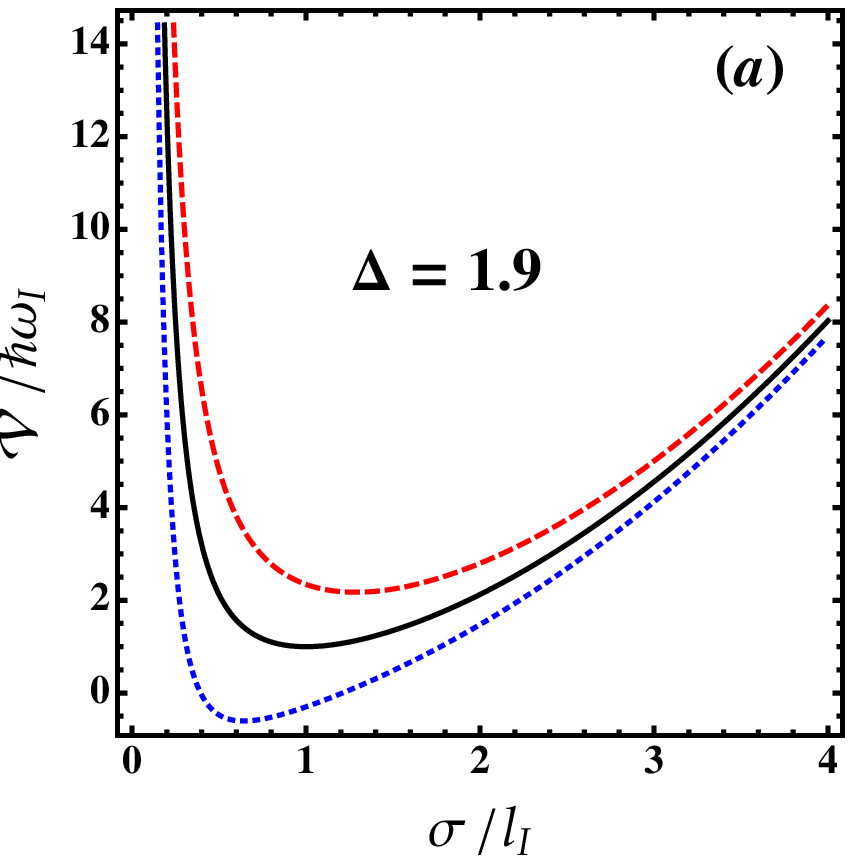} 
\includegraphics[scale=0.45]{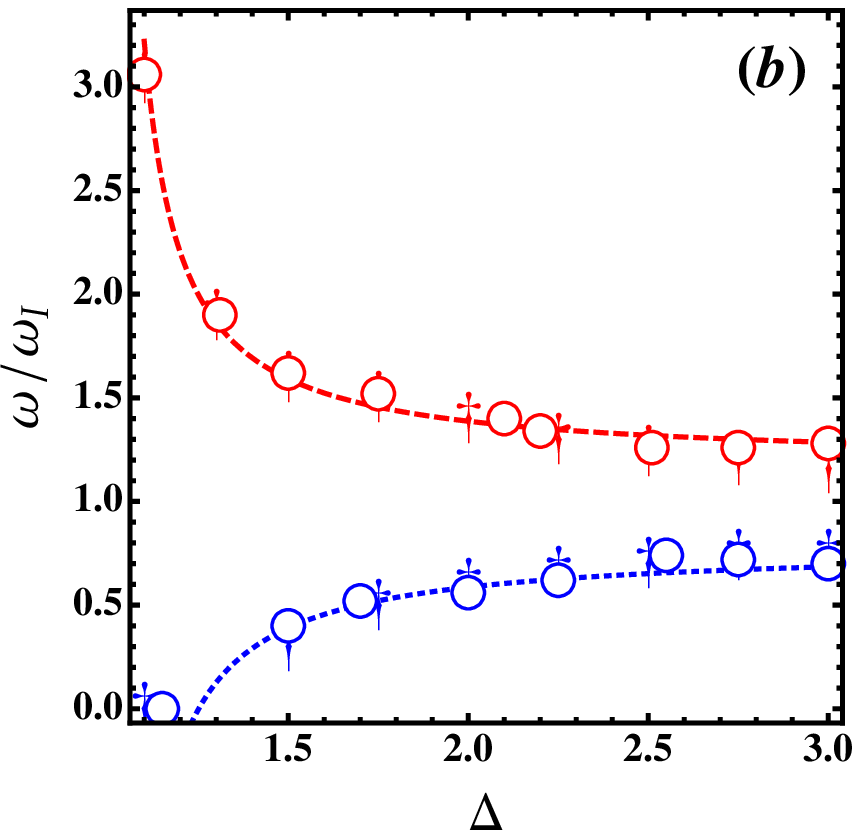} 
  \caption{ Homogeneous mixture. (a) The effective potential from Eq.(\ref{pot1})  for several values of $\gamma_1$ and $\gamma_2$.
Black solid lines: $\gamma_1=\gamma_2=0$. Red dashed lines: $\gamma_1=0.5$ and $\gamma_2=3$. Blue dotted lines:  $\gamma_1=3$ and $\gamma_2=0.5$.
(b) The frequency of the breathing oscillations $\omega/\omega_I$ as a function of $\Delta$ for a miscible mixture.
Dashed and dotted lines: variational calculations, $\omega= \sqrt{\partial^2  {\cal V}/ \partial \sigma^2}|_{\sigma=\sigma_0}$.
Crosses: numerical simulations of Eq.(\ref{GPE}).  Open circles: numerical simulations of coupled BdG equations.
%Numerically, we compute the low-lying excitations to a sudden small change of the impurities trap frequencies : $\omega_I= (1+\epsilon) \omega_I$, where $\epsilon=0.005$.
Here $l_I=\sqrt{\hbar/m_I\omega_I} $ is the impurities oscillator length. } 
    \label{HPF}
 \end{center}  
\end{figure}

%shows that for $\gamma_j=0$, the effective potential $ {\cal V}$ is fairly independent of the miscibility of the two fluids.
Figure.\ref{HPF}.a depicts that the potential $ {\cal V}$ develops a local minimum at $\sigma=\sigma_0$  revealing the localization of the impurities
in the ground state.
The position and the depth of such a minimum depend on the BEC-impurity interactions. 
For relatively large $\gamma_j $,  the potential ${\cal V}$ develops a deep minimum signaling that the impurities are strongly localized.
For instance, for $\gamma_1=3$ and $\gamma_2=0.5$,  the potential has a local minimum at $\sigma_0 \simeq 0.6 l_I$.  
An important remark is that the effective potential is sensitive to switching relative strengths of interaction  $\gamma_j$.
This fact can be straightforwardly interpreted: the self-trapping term, $- 4 g_{Ij} N_I^{1/2}(\gamma_j- \gamma_{3-j} g_{12}/g_j )$, 
which for fixed value of $g_{12}/g_j$, imparts an additional attractive/repulsive force on the effective potential shifting 
the local minimum either in an upward (for $\gamma_j < \gamma_{3-j} g_{12}/g_j$) or downward (for $\gamma_j> \gamma_{3-j} g_{12}/g_j$)
direction with respect to the $\gamma_j=0$ case.
The interplay of the intra- and interspecies interactions ($g_j$, $g_{12}$), may also affect the position of the local minimum 
and hence, the localization of the impurities. 

Figure \ref{HPF}.b. shows that for $\gamma_1> \gamma_2$ i.e the BEC1-impurity interaction is stronger than the BEC2-impurity interaction, 
the self-trapping term provides an extra repulsive force results in the frequency of the breathing oscillations of impurities $\omega_I$ inside the miscible environment
increases with the miscibility parameter $\Delta$.
Whereas,  for $\gamma_1<\gamma_2$, the frequency of oscillations $\omega_I$ lowers with $\Delta$ since the self-trapping term becomes negative.
The variational results and those obtained from the numerical simulation of Eq.(\ref{GPE}) and of the coupled GP equations (\ref{T:DH1})-(\ref{T:DH2})
are in good agreement.

%This fact has a physical explanation. The  self-trapping term $-\Delta/(\Delta-1)  g_{IBj} N_I^{1/2} (- \gamma_{3-j} g_{B12}/g_{Bj})$
%imparts an additional attractive force on the impurity's effective potential shifting both the height and width of the local mimimum results in 
%a robust localization of the impurity (see Fig.\ref{HPF}.c). 

%Whereas, if the two Bose components are phase separate, strong localization occurs only at at a small $\gamma_1$ and large $\gamma_2$ 
%($\sigma_0 = 0.2 l_0$ for $\gamma_1=0.5$ and large $\gamma_2=3$). 
%This fact has a physical explanation. The  self-trapping term $-\Delta/(\Delta-1)  g_{IBj} N_I^{1/2} (- \gamma_{3-j} g_{B12}/g_{Bj})$
%imparts an additional attractive force on the impurity's effective potential shifting both the height and width of the local mimimum results in 
%a robust localization of the impurity (see Fig.\ref{HPF}.c). 

\subsection {Trapped mixture}

\begin{figure}
\begin{center}
\includegraphics[scale=0.45]{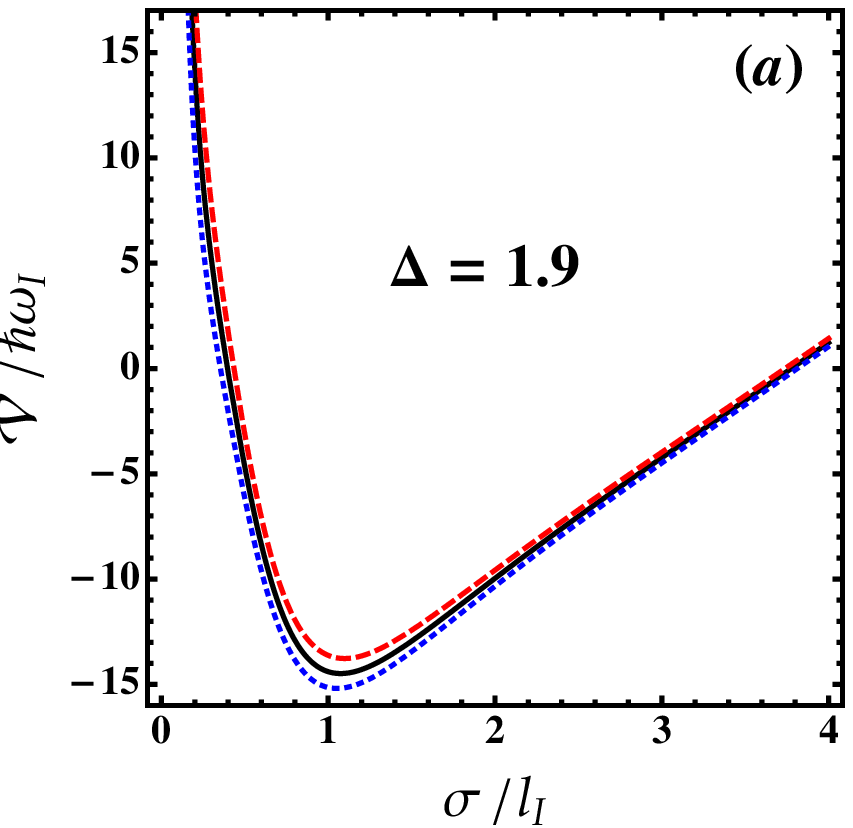} 
\includegraphics[scale=0.45]{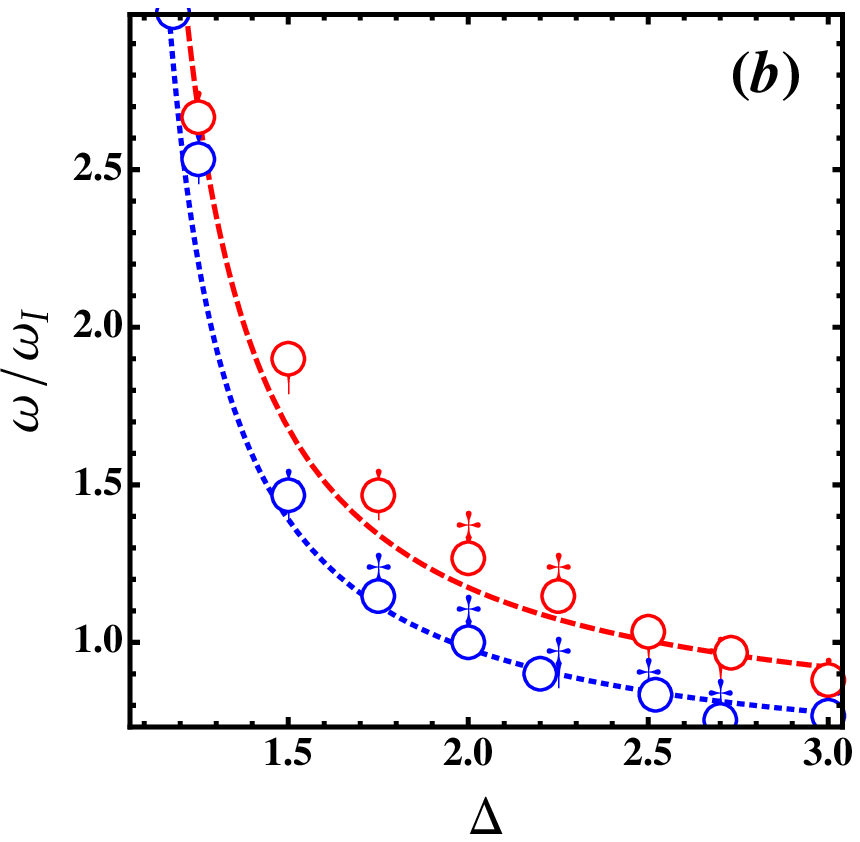} 
  \caption{ The same as Fig.\ref{HPF} but for impurities immersed in a trapped Bose-Bose mixture. 
The harmonic frequencies are given as:  $\omega_{j \perp}=2\pi \times 34$ kHz,  
$\omega_{j x}=2\pi \times 62$ Hz \cite{Cat1}, $\omega_{I \perp}=2\pi \times 50$ kHz,  and $\omega_{I x}=2\pi \times 90$ Hz. 
Black solid lines: $\gamma_1=\gamma_2=0$. Red dashed lines: $\gamma_1=0.5$ and $\gamma_2=3$. Blue dotted lines:  $\gamma_1=3$ and $\gamma_2=0.5$.}
    \label{TPF}
 \end{center}  
\end{figure}

The trap may modify the properties of the impurities and the self-trapping process.
The behavior of the total effective potential of Eq.(\ref{pot1}) is captured in Fig.\ref{TPF}.a.
We see that ${\cal V}$ has a minimum at $\sigma_0 \gtrsim l_I$ whatever the values of $\gamma_j$. 
It is flattened near its minimum leading to broaden the width of the impurity.
This is most likely due to the inhomogeneity in the mixture arising from the external trapping force.
Note that a similar behavior occurs in a single BEC interacting with an impurity \cite{Cat1, THJ}.
Furthermore, the BEC-impurity couplings $\gamma_1$ and $\gamma_2$ appear to bring opposite effects on the effective potential 
as in the homogeneous case. This property is clearly visible in Fig.\ref{TPF}.a.

Figure \ref{TPF}.b depicts that the frequency of the breathing modes are decreasing with $\Delta$ for any $\gamma_j$ which
is in stark contrast with the homogeneous case where $\omega_I$ varies in the opposite way with $\Delta$.
The diminution of the oscillations frequency can be likely interpreted by the fact that they are dominated by the harmonic frequency.
Our numerical and analytical findings show excellent agreement with each other in the whole range of the excitations spectrum.

\section{Dynamics of impurities in a homogeneous miscible bath} \label{DIH}

\begin{figure}
\begin{center}
\includegraphics[scale=0.9]{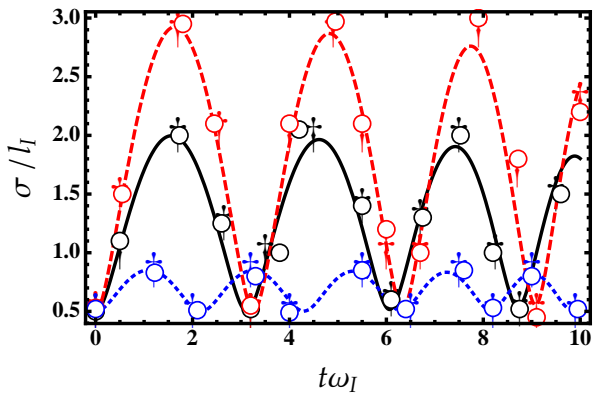} 
  \caption{Dynamic evolution of trapped impurities axial width in homogeneous miscible Bose mixtures 
for different relative interaction strengths $\gamma_j$.
Parameters are the same as in Fig.\ref{HPF}. 
Dashed lines: Variational results of Eq.(\ref{Veq2}). Crosses: numerical simulations of Eq.(\ref{GPE}).  
Open circles: numerical simulations of coupled GP equations (\ref{T:DH1})-(\ref{T:DH2}).}
    \label{dyn}
 \end{center}  
\end{figure}

In this section we will shed some light on the dynamics of few impurities interacting with homogeneous miscible bath.
We compute the time evolution of the impurities axial width after a sudden decrease in $\omega_I$ in the weak coupling regime by solving the variational equation (\ref{Veq2}).  
Figure.\ref{dyn} shows that the oscillation amplitude and width of $\sigma$ are increasing with $\gamma_2$  and decreasing with $\gamma_1$.
The impurities oscillate faster for $\gamma_1>\gamma_2$.
This is in contrast to  the case of polarons in a single component BEC where the oscillation amplitude of $\sigma$ found to be large (small) in the absence (presence) of 
the BEC-impurity interaction \cite{Cat1,THJ}. 
To check the obtained variational solutions, we numerically solve the extended self-focussing NLSE (\ref{GPE}) and the coupled GP equations (\ref{T:DH1})-(\ref{T:DH2})
using the time-splitting method \cite{THJ}. 
We find that the analytical findings display excellent agreement with our numerical solutions at times $t \omega_I \leq 7$,
while at larger times there is a distinct difference between the variational and numerical calculations. 
Numerically, the width can be calculated through $\sigma(t)=\sqrt{\langle x^2 \rangle}$.

%The situation is inverted in the immiscible case where very fast vibrations with small amplitudes are observed for large $\gamma_1$ (see Fig.\ref{dyn}.b)
%which are justified by the fact that the self-trapping term is negligible compared the kinetic energy.
%Whereas for small $\gamma_1$, the  oscillations amplitude become $\sim 2$ times greater than those perceived in the miscible bath.
%In the absence of an impurity-condenstate interaction, the oscillation amplitude and size of $\sigma$ remain unchanged by passing from a miscible to an immiscible mixture

For $\gamma_j=0$, one can expect that impurities simply oscillate back and forth with oscillation frequency, 
that is exactly the frequency of the axial harmonic confinement for impurity atoms.
Similar behavior holds true in a single BEC-impurity mixture \cite{Guebli, Ling}.

More interestingly, Fig.\ref{dyn} shows that the breathing oscillations are damped out. 
The damping occurs most probably due to the interaction of impurities with Bogoliubov phonons that
are created during the oscillations of the impurities. 
This can be checked easily by means of the standard Bogoliubov method.
Working in momentum space, the Hamiltonian of the mixture takes the form:
$\hat H= E_0+\sum_{j=1}^2\big[ \sum_{\bf k} \hbar \omega_{kj} \hat b_{{\bf k} j} ^\dagger \hat b_{{\bf k} j} + g_{Ij} n_j+ g_{Ij} \sum_{\bf k\neq 0} 
(\hat b_{{\bf k} j}^\dagger+ \hat b_{{\bf k} j}) f_{{\bf k} j} \big]$,
where $E_0$ is the ground state energy, the Bogoliubov frequencies $\omega_{kj}$ can be calculated from the above BdG equations.
In the phonon regime they can be written as $\omega_{k \pm}=\hbar c_{\pm} k$, where sound velocities corresponding to the upper $c_+$  and lower $c_-$ branches are
$c_{\pm} ^2=[ c_1^2+c_2^2 \pm \sqrt{ ( c_1^2-c_2^2) ^2 + 4 \Delta^{-1} c_1^2 c_2^2} ]/2$ 
with $c_j=\sqrt{\mu_j/m_j}$ being the sound velocity of each component.
The function $f_{{\bf k} j} =\sqrt{n_j E_{kj}/ V \hbar \omega_{kj} } \int d x  |\Phi_I (x)|^2\exp ({i.k.x})$ 
depends parametrically on the oscillating width $\sigma$  \cite{Cat1,THJ}, where $\Phi_I (x)$ is defined in Eq.(\ref{Gaus}).
The interaction of impurities with their upper branch phonon environment results in dissipation in energy which is responsible for the damping similarly to the single polaron case \cite{THJ}.
The lower branch of the spectrum is unstable, thus the interaction of impurities with the two BEC may lead to a dynamic oscillatory instability of the impurity state.

\section{Conclusions} \label{concl}

In this paper we theoretically investigated, for the first time to our knowledge, the static and the dynamics of 
quasi-1D repulsive polarons in Bose-Bose mixtures at zero temperature. The impurities are assumed to be pinned in the center of the trap.
Within the mean-field theory we derived  self-consistent three coupled differential equations for the two condensates and the impurities wavefunctions.
Effects of the impurities on the miscibility of the mixture has been studied in details by numerically solving such equations in the static case.
The outcomes of this simulation revealed that the impurities lead to modify the shape of the two condensates and the binding energy of the mixture polarons.
We found that at certain BEC-impurity interaction strengths, the mixture undergoes miscible-immiscible phase transition.   
In such a transition the total depletion is lowered  and the impurities becomes markadly bound with their bosonic mixture host.
This important feature has never been observed before in the literature. 

On the other hand, we studied the breathing oscillations and the time evolution of a harmonically trapped impurity in homogeneous and inhomogeneous dual BECs 
in the framework of the TF regime. The validity criterion of this approach has been accurately established.
We derived a self-focussing nonlinear equation describing the time evolution of such a system. 
Employing a suitable variational ansatz,  we computed the effective potential and the breathing oscillations of the impurities in terms of the miscibility parameter
for both homogeneous and inhomogeneous baths. 
Our results pointed out that the BEC-impurity interactions, the trapping force and the miscibility parameter may strongly affect
the localization process and the breathing oscillations frequencies of the impurities.  
For instance, in the case of impurities immersed in a homogeneous Bose mixture, we pointed out that
upon swapping the values of $\gamma_1$ and $\gamma_2$, the breathing frequency varies in the opposite way with the miscibility parameter.
Whereas, in the inhomogeneous case the impurities oscillations are continuously decreasing with $\Delta$ for any $\gamma_j$ due to the external harmonic force.
The variational frequencies agree quite well with those obtained from the BdG equations.

Moreover, we deeply analyzed the time evolution of the impurities width.  
We found that the breathing modes of the width display a strong dependence  on the relative coupling strengths $\gamma_j$ and on the miscibility parameter $\Delta$.
Their amplitudes are reduced for larger times due to the phonon-impurity, BEC-BEC and BEC-impurity interactions.
Our analytical results have been checked  through comparison with a direct numerical simulation of coupled GP equations.

Experimentally, the existence of polarons in Bose-Bose mixtures can be demonstrated by using radio frequency spectroscopy technique,
utilized for single Bose polarons (see e.g.\cite{Jor}). 
Our results open new avenues to investigate impurity transport in multicomponent Bose systems.  
They offer fascinating prospects for exploring new exotic molecular bound states due to the coexistence of several sorts of coupling interactions.
One should stress that the present work can be readily extended to the case of attractive Bose-impurity couplings.
In future work we will attempt to highlight the role impurities play in mixture droplets \cite{Petrov, Cab, Semg, Errico,Boudj12, Boudj00}.

%To summarize, there are three major results from our analysis of variance

%\section{Acknowledgements}
%We are indebted to ... for stimulating discussions.

\end{document}